\documentclass[twocolumn,aps,prl,amsmath,superscriptaddress,notitlepage,longbibliography,nobibnotes]{revtex4-2}
\usepackage{amssymb}
\usepackage{amsmath}
\usepackage{mathrsfs}
\usepackage{graphicx}
\usepackage{float}
\usepackage[normalem]{ulem}
\usepackage{color}
\usepackage{bm}

\begin{document}
\title{Hybrid skin-topological modes without asymmetric couplings}
\author{Weiwei Zhu}
\email{phyzhuw@gmail.com}
\affiliation{Department of Physics, National University of Singapore, 117542, Singapore}
\author{Jiangbin Gong}
\email{phygj@nus.edu.sg}
\affiliation{Department of Physics, National University of Singapore, 117542, Singapore}

\pagebreak

\begin{abstract}

Abstract:
Non-Hermitian skin effect (NHSE) in non-Hermitian lattice systems, associated with a point gap on the complex energy plane,  has attracted great theoretical and experimental interest.  Much less is studied on the so-called second-order non-Hermitian skin effect, where the bulk does not support a point gap but localization at the corner still occurs.  This work discovers a class of hybrid skin-topological modes as the second-order non-Hermitian skin effect without asymmetric couplings.  Specifically, by only adding gain/loss to two-dimensional Chern insulators and so long as the gain/loss strength does not close the line gap, all the topological edge states are localized at one corner under the open boundary condition, with the bulk states extended.   The resultant non-Hermitian Chern bands can be still topologically characterized by Chern numbers, whereas the hybrid skin-topological modes are understood via some auxiliary Hermitian systems that belong to either intrinsic or extrinsic second-order topological insulator phases. By proposing an innovative construction of auxiliary Hamiltonian,  our generic route to hybrid skin-topological modes is further successfully extended to nonequilibrium topological systems with gain and loss, where the anomalous Floquet band topology is no longer captured by band Chern numbers.   The extension thus leads to the intriguing finding of nonequilibrium hybrid skin-topological modes. In addition to offering a straightforward route to experimental realization of hybrid topological-skin effects,  this study also opens up a promising perspective for the understanding of corner localization by revealing the synergy of three important concepts, namely, non-Hermitian topological insulator, second-order non-Hermitian skin effect, and second-order topological insulator.

\bigskip

\end{abstract}

\maketitle

\section{Introduction}
Non-Hermitian systems are now widely known to support topological states that may not exist in Hermitian system~\cite{El-Ganainy2018,Ashida2021,Bergholtz2021}.  The bulk spectrum of non-Hermitian lattice systems under the periodic boundary condition (PBC) is typically complex, and the gap can be either a line gap or a point gap. Different gaps associated with specific symmetries can then be used to classify a wide variety of non-Hermitian topological phases~\cite{Gong2018,Kawabata2019}.
Though line-gap topology may be transformed to a Hermitian counterpart without closing the band gap,  point-gap topology is unique to non-Hermitian systems.  In particular, point-gap topology is the  underlying topological protection responsible for the so-called non-Hermitian skin effect (NHSE) ~\cite{Okuma2020,Zhang2020}, causing all the bulk states to be localized under the open boundary condition (OBC)~\cite{Yao2018,Kunst2018,McDonald2018,Kunst2019,Longhi2019,Jin2019,Song2019,Lee2019a,Borgnia2020,Lee2019,Deng2019,Jiang2019,Li2020}.

NHSE, as guaranteed by a point-gap topology, has spurred a great deal of theoretical interest because it breaks the usual bulk-edge correspondence for line-gap topology~\cite{Ghatak2019,Mochizuki2019}. Indeed, non-Bloch topological band theory has been developed to recover the bulk-edge correspondence for non-Hermitian systems~\cite{Yao2018,Yao2018a,Yokomizo2019,Imura2019,Kawabata2020,Longhi2020,Scheibner2020,Zhang2020a,Yang2020,Zhu2020,Ghatak2020,Helbig2020,Xiao2020,Zhu2021,Zhou2021,Wu2021}.
A variety of physical platforms, such as  cold atoms, photonics, electrical circuits, mechanics, and acoustics have been developed to study NHSE experimentally, with motivating findings~\cite{Brandenbourger2019,Hofmann2020,Weidemann2020,Liu2021,Zhang2021,Chen2021,Palacios2021,Zhang2021a,Zou2021,Zhang2021b,Liang2022}.

It is fair to say that most studies to date have focused on NHSE in one-dimensional (1D) systems.   In other case studies that do involve two-dimensional (2D) systems, the interplay of NHSE and topological bands is already found to bring interesting physics, such as non-Hermitian Chern bands~\cite{Yao2018a,Kawabata2018,Chen2018,Hirsbrunner2019,Wu2019,Xiao2022},  defect induced NHSE~\cite{Sun2021,Schindler2021,Bhargava2021,Manna2022}, and  the hybrid skin-topological modes~\cite{Lee2019,Li2020a,Zou2021,Shang2022} where corner localization is due to the interplay of topological localization and NHSE.  Nevertheless, as far as NHSE is concerned in these studies of 2D systems, it is essentially the 1D (or first-order) NHSE that is in force, because the source of NHSE can be clearly attributed to one direction, but not the other direction.  In particular,
to date our knowledge of hybrid skin-topological mode is based on a specific lattice design~\cite{Lee2019,Li2020a}, namely, the manifestation of asymmetric coupling in one direction in the presence of topological localization in the other direction.

There have also been growing interest in the so-called second-order NHSE that can localize certain states at the corner while all the bulk states are still extended  under OBC~\cite{Kawabata2020a,Okugawa2020,Fu2021,Kim2021}.  Second-order NHSE is remarkable because the bulk band does not even support a point gap, so the resulting skin corner models are clearly beyond the usual paradigm of NHSE.   A point gap now emerges only after one of the dimensions is placed under OBC and the second dimension is under PBC.   This being the case,  skin corner modes can be obtained after placing two different dimensions under OBC, hence of second-order.    Such type of skin corner modes is of second-order for another physical reason.  That is, available studies \cite{Kawabata2020a,Okugawa2020,Fu2021,Kim2021} indicate that second-order NHSE can be connected with a second-order topological insulator phase of some auxiliary Hamiltonian.   However, all theoretical models proposed to date require asymmetric couplings (plus,  positive couplings along one direction and negative couplings along anther direction).  These constructions are convenient in theory, but far from trivial in experimental realizations.

Echoing with the possibility of inducing  non-Hermitian topological phase transitions solely by gain and loss~\cite{Takata2018,Luo2019,Ao2020,Song2020,Gao2021,Weidemann2022},  in this work we shall reveal a generic route to hybrid skin-topological modes without asymmetric couplings,  by applying solely gain and loss to topological insulators with chiral edge states, for both conventional Chern insulator phases and periodically driven (hence nonequilibrium) topological phases.  The subtle interplay between topology and non-Hermitian effects then leads to hybrid skin-topological modes, which can be identified as a new type of second-order NHSE. In particular, upon the introduction of some gain and loss that still maintains  the line-gap topology of the starting topological insulator phase, the system continues to support chiral edge states, which now however acquire gain and loss as they propagate along the edges of the system. For example, in the case of a non-Hermitian Chern insulator, it is still topologically described by a non-zero Chern number upon introducing gain and loss.   Remarkably,  we show that  skin corner modes generically exist under OBC, with their number being proportional to the length of a 2D system and  localized at one corner of the system only.  As shown below, the localization behavior of the obtained corner skin models can be phenomenologically understood by state accumulation of chiral edge states with gain and loss.  Furthermore, the existence of corner skin models can also be understood through the construction of an auxiliary Hamiltonian~\cite{Benalcazar2017,Peterson2018,Serra-Garcia2018,Imhof2018,Mittal2019,Xie2019,Chen2019,Liu2019,Xue2018,Ni2018,Zhang2019,Zhu2021a,Li2021,Niu2021,Zhu2022}.
That is, the non-Hermitian systems under consideration must host skin corner modes if the auxiliary Hamiltonian is in a second-order topological insulator phase, protected by a bulk gap (intrinsic second-order) or an edge-state gap inside the bulk gap (extrinsic second-order)~\cite{Geier2018}. { In either case the introduced gain and loss can induce an edge band gap opening and thus a phase transition to the second-order topological insulating phase. We also show that the nature of hybrid skin-topological modes allows for the construction of a topological switch to turn on/off skin effects, simply by introducing topological phase transitions.}

As compared with the asymmetric couplings previously assumed in second-order NHSE or hybrid skin-topological modes~\cite{Lee2019,Li2020a,Zou2021,Shang2022},  the physics revealed in this work is widely applicable because gain or loss is much easier to realize.
As a matter of fact, loss is ubiquitous due to absorbing materials or some leaky modes, and gain can also be introduced by optical or electrical pumping in photonic systems.  Because our route to hybrid skin-topological modes is rather general, we are motivated to explore the possibility of nonequilibrium skin-topological modes.  We indeed manage to extend our approach to nonequilibrium topological systems with gain and loss, where the anomalous Floquet band topology is no longer captured by band Chern numbers.  To better connect nonequilibrium hybrid skin-topological modes with second-order topology, we have proposed an innovative construction of auxiliary Hamiltonian to treat periodically driven topological systems with gain and loss.


\section{Hybrid skin-topological modes in a non-Hermitian Haldane model}
 \subsection{Haldane model with gain and loss}
\begin{figure}[htbp]
\includegraphics[width=1\linewidth]{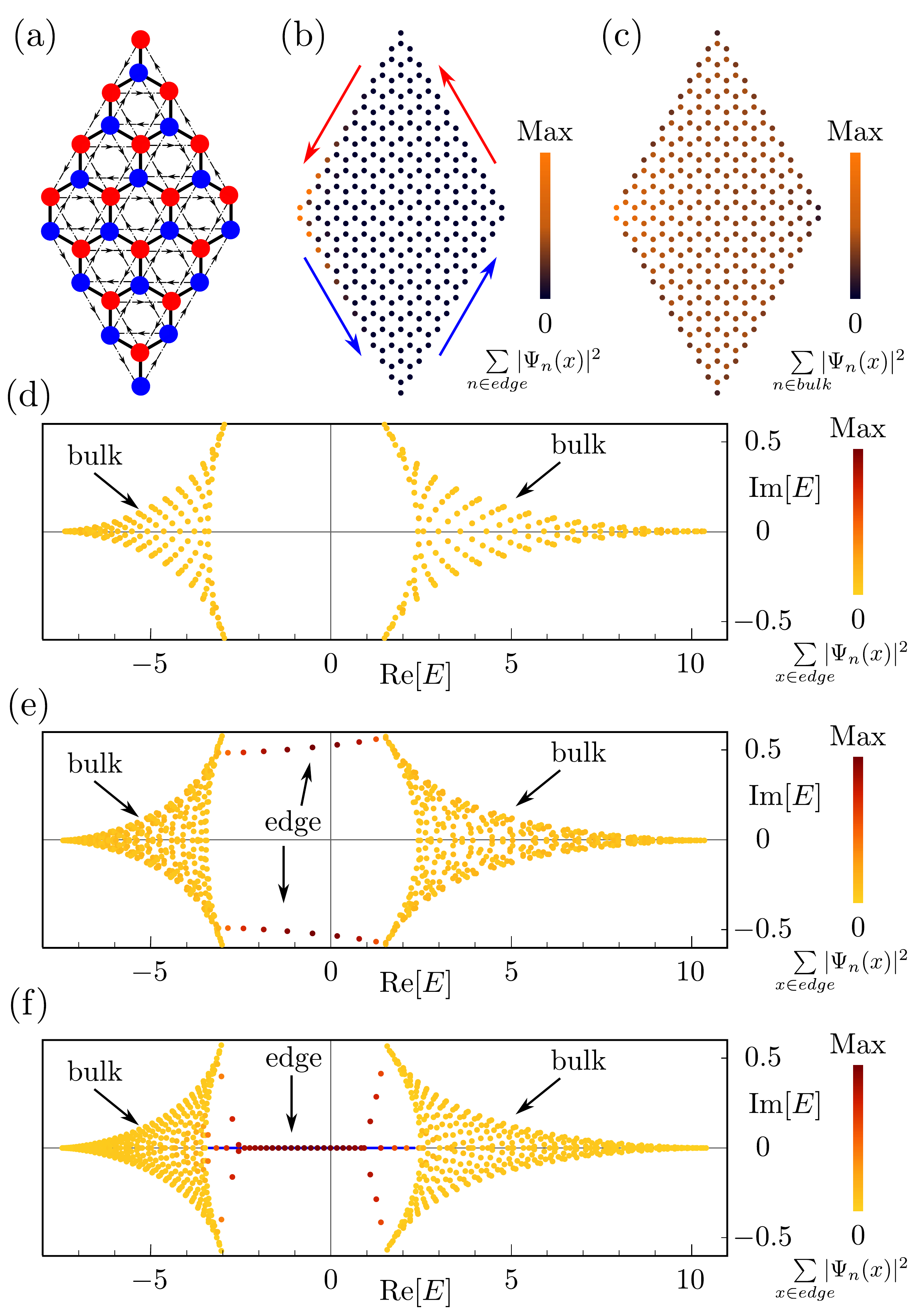}
\caption{(Color online)
Hybrid skin-topological modes as a second-order non-Hermitian skin effect in a non-Hermitian Haldane model. (a) Non-Hermitian Haldane model with gain (blue) and loss (red).
(b) Summation of the state density profile over all edge states $\sum_{n\in edge}|\Psi_n(x)|^2$. Color represents the strength of localization on the left edge $\sum_{x\in edge}|\Psi_n(x)|^2$.
(c) Summation of the state density profile over all bulk states $\sum_{n\in bulk}|\Psi_n(x)|^2$. Color represents the strength of localization of all the bulk states $\sum_{x\in bulk}|\Psi_n(x)|^2$.
(d) Spectrum of the system under PBC along both directions. Color represents the strength of localization on the left edge $\sum_{x\in edge}|\Psi_n(x)|^2$
(e) Spectrum of a semi-infinite structure with PBC along $x$ and OBC along $y$.  Color represents the strength of localization on the left edge $\sum_{x\in edge}|\Psi_n(x)|^2$.
(f) Spectrum of a finite structure with OBC along both directions.   Color represents the strength of localization on the left edge $\sum_{x\in edge}|\Psi_n(x)|^2$.
}
\label{Haldane}
\end{figure}

We start with a non-Hermitian Haldane model,  as depicted in Fig.~\ref{Haldane}(a).  This non-Hermitian version of the Haldane model is obtained by introducing gain and loss to two sublattices respectively~\cite{Haldane1988}. The resulting  tight-binding lattice Hamiltonian can be expressed as follows:
\begin{equation}\label{eq1}
\begin{split}
  H= &t_1\sum\limits_{\langle i,j\rangle}c_i^\dag c_j+t_2 e^{i\nu_{ij}\phi}\sum\limits_{\langle\langle i,j\rangle\rangle}c_i^\dag c_j\\
  &+ig\sum\limits_{i\in A}c_i^\dag c_i-ig\sum\limits_{i\in B}c_i^\dag c_i\\
\end{split}
\end{equation}
where $c_i^\dag$ ($c_i$) is the creation (annihilation) operator for a particle at the $i$th site.
The first term is the nearest-neighbor hopping with an amplitude $t_1$.
The second term is the next-nearest-neighbor hopping with an amplitude $t_2$ and a phase $\nu_{ij}\phi$. The phase is direction dependent. If the hopping is along the arrows in Fig.~\ref{Haldane}(a), the phase is positive $\nu_{ij}=+1$. If the hopping is in the opposite direction, the phase is negative $\nu_{ij}=-1$. The second term breaks the time reversal symmetry so that the system can support quantum anomalous Hall effect.
The third term is what we introduce in this work, representing the on-site gain (loss) on sublattice A (sublattice B).   In the absence of
such gain and loss, the system is a well-known Chern insulator and supports topological chiral edge states.

As a rather standard treatment, one can now apply a Fourier transformation to the real-space Hamiltonian in Eq.~\ref{eq1},  obtaining the momentum-space Hamiltonian:
\begin{equation}\label{eq2}
 H(\mathbf{k})=\vec{d}(\mathbf{k})\cdot\vec{\sigma}+ig\sigma_z
\end{equation}
where $\vec{d}(\mathbf{k})=(d_0(\mathbf{k}),d_x(\mathbf{k}),d_y(\mathbf{k}),d_z(\mathbf{k}))$ is a vector as function of $\mathbf{k}$, $\vec{\sigma}=(\sigma_0,\sigma_x,\sigma_y,\sigma_z)$ and
\begin{equation}\label{eq3}
\begin{split}
  d_0(\mathbf{k})=&2t_2\cos\phi[\cos k_x+2\cos(k_x/2)\cos(\sqrt{3}k_y/2)],\\
  d_x(\mathbf{k})=&t_1[1+2\cos(k_x/2)\cos(\sqrt{3}k_y/2)],\\
  d_y(\mathbf{k})=&2t_1\cos(k_x/2)\sin(\sqrt{3}k_y/2),\\
  d_z(\mathbf{k})=&-2t_2\sin\phi[\sin k_x+2\sin(k_x/2)\cos(\sqrt{3}k_y/2)],\\
\end{split}
\end{equation}
with $\mathbf{k}=(k_x,k_y)$ being quasimomentum, $\sigma_0$ being two-by-two identity matrix and $\sigma_i$, $i=x,y,z$ being the Pauli matrices.

Let us first investigate the spectrum of the above-described non-Hermitian Haldane model under the open boundary condition (OBC).
In our sample calculations, we use the following system parameters $t_1=3$, $t_2=0.5$, $g=0.6$ and $\phi=\pi/3$ unless specified otherwise.
For zero gain and loss, the model system here is a Chern insulator with chiral edge states propagating along the system's edge with a definite direction due to time-reversal-symmetry breaking ~\cite{Haldane1988}.  Upon introduction of the gain and loss to two sublattices, it is expected that there will be no band gap closure  so long as the strength of the gain and loss terms is not too strong.   Remarkably, all the identified topological edge states in the bulk gap are found to be localized at the system's left corner, as shown in Fig.~\ref{Haldane}(b) whereas all the bulk states are extended as shown in Fig.~\ref{Haldane}(c).  Furthermore, without a line-gap topological transition, the system in the presence of the gain and loss is still classified as a Chern insulator,  but as a non-Hermitian version.
Fig.~\ref{Haldane}(d)-(f) compare the spectrum plotted on the complex energy plane under PBC, mixed PBC-OBC, and OBC.
The PBC spectrum in Fig.~\ref{Haldane}(d) depicts two well-separated bulk bands.   The spectrum of the system of an semi-infinite (strip) structure in Fig.~\ref{Haldane}(e) shows that the chiral edge states together with the bulk spectrum encloses a nonzero area, thus potentially allowing a point gap.
In Fig.~\ref{Haldane}(f) for the system under OBC,  two well-separated bulk bands are seen.  Between the two bulk bands in Fig.~\ref{Haldane}(f),  there are gapless topological edge states connecting the two bulk gaps.   That  the energies of the chiral edge states here are dramatically different when we change the boundary condition from mixed PBC-OBC to OBC strongly suggests that the observed corner modes here are due to the skin localization of chiral edge states and hence represent hybrid skin-topological modes.

\begin{figure}[htbp]
\includegraphics[width=1\linewidth]{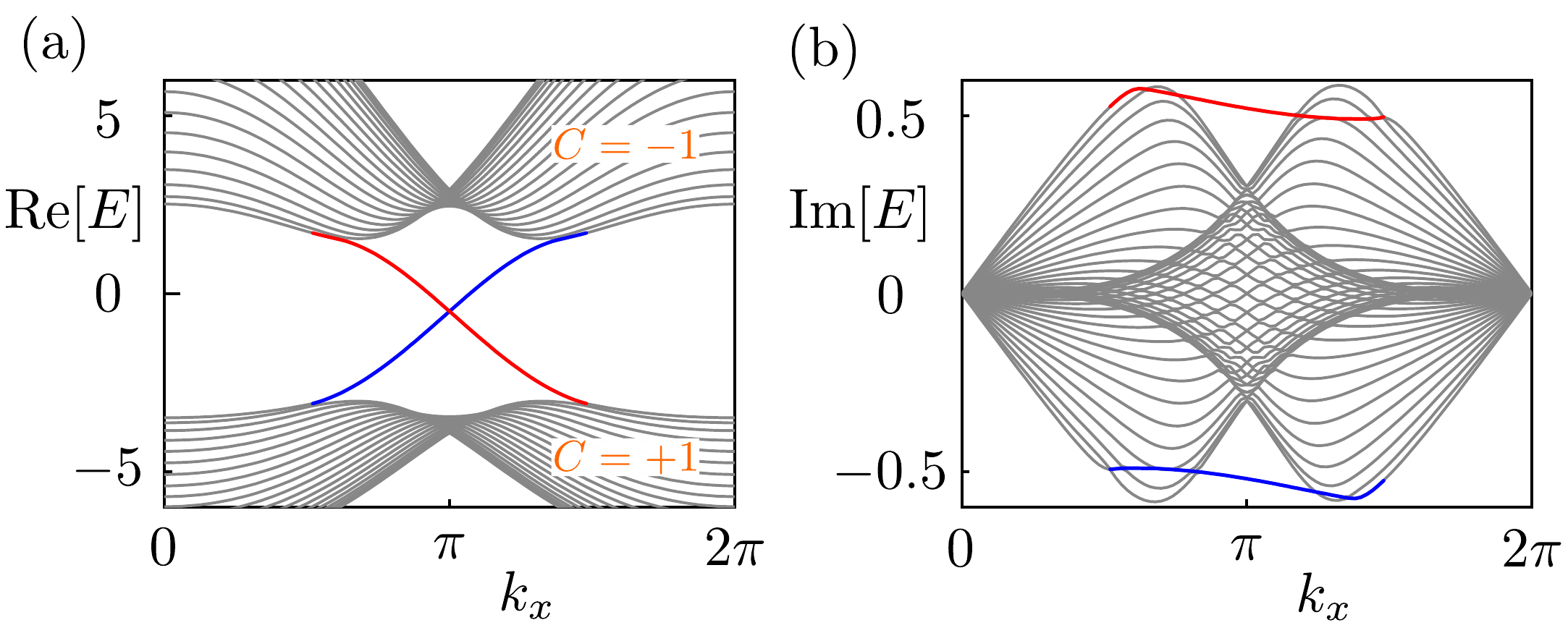}
\caption{(Color online) Topological edge states of a non-Hermitian Haldane model in a zigzag strip structure. (a) Real part of the energy bands. (b) Imaginary part of the energy bands.    Note that the edge states propagate to the left has positive imaginary parts and hence acquire gain, and the edge states propagate to the right has negative imaginary parts and hence experience loss.}
\label{zigzag}
\end{figure}

It is now necessary to digest why the topological chiral edge states are all localized at one corner of the system. To that end we next look into more details of the energy bands of this system in a zigzag strip structure with periodic boundary condition (PBC) along $x$ and  OBC along $y$.
The real part and imaginary parts of the complex spectrum under such a mixed PBC-OBC boundary condition are shown in Fig.~\ref{zigzag}(a) and Fig.~\ref{zigzag}(b), respectively, as a function of the Bloch momentum $k_x$ along the $x$ direction.  From Fig.~\ref{zigzag}(a) and Fig.~\ref{zigzag}(b) one clearly sees that
there are two topological edge channels marked by red color and blue color. The red one represents the edge channel propagating towards the left direction with gain (because of the positive imaginary part), whereas the blue one represents the edge channel that only allows transport to the right with loss (because of the negative imaginary part).  To reflect this observation,  in Fig.~\ref{Haldane}(b) we further mark the gain edge states and loss edge states by blue arrows and red arrows.  Evidently, then, as gain edge states propagate to the left with gain, it tends to accumulate population on the left corner.   Interestingly and analogously, because the loss edge state propagate to the right with loss,  the loss edge states prefer to accumulate to the left corner.  Thus, no matter what edge state channels we inspect, they all tend to be localized at the left corner.  This is a noteworthy feature of the system and can only be taken as a phenomenological explanation of why skin corner modes should generically emerge in a non-Hermitian Chern insulator.

The non-Hermitian Haldane model thus depicts a non-Hermitian Chern insulator, of which the topological characterization can be captured by
the following Chern number defined as~\cite{Yao2018a}
\begin{equation}\label{eq4}
  C_n=\frac{1}{2\pi i}\int_{BZ}d^2\mathbf{k}\epsilon_{ij}\langle\partial_iu_{Ln}(\mathbf{k})|\partial_ju_{Rn}(\mathbf{k})\rangle
\end{equation}
where $n$ is the band number, $i,j=x,y$ and $\epsilon_{xy}=-\epsilon_{yx}=1$. $|u_{Ln}(\mathbf{k})\rangle$ and $|u_{Rn}(\mathbf{k})\rangle$ are left and right eigenvectors of $H(\mathbf{k})$ with normalization condition $\langle u_{Ln}(\mathbf{k})|u_{Rn}(\mathbf{k})\rangle=1$.
Because all the bulk bands here are extended,  the conventional (first-order) NHSE is not manifested by the bulk bands under PBC and hence the non-Hermitian  band Chern numbers can be defined in the usual Brillouin zone, without invoking the so-called non-Bloch Chern number in the generalized Brillouin zone.
Indeed, the Chern numbers for the two bands based on  the usual Brillouin zone  are $\pm1$, as shown in Fig.~\ref{zigzag}(a).
The conventional bulk-boundary correspondence is also obeyed: there are two chiral edge channels in the band gap, as shown in  Fig.~\ref{zigzag}, propagating towards opposite directions and found at opposite edges of the strip geometry.


\subsection{Topological characterization of hybrid skin-topological modes}
The topological Chern numbers discussed at the end of the previous section can be used to understand the bulk-edge correspondence of the non-Hermitian Chern insular in a strip geometry.   However, it remains to establish the bulk-corner correspondence in order to reveal the nature of hybrid skin-topological modes discovered above.
For conventional (first-order) NHSE, it is known to have a  topological origin and can be associated with some nonzero winding number of the spectrum on the complex plane~\cite{Okuma2020,Zhang2020}.   By analogy, the skin corner modes arising from the localization of chiral edge modes can be associated with some nonzero spectral winding number of non-Hermitian chiral edge states.  However, in practice it is not an easy task to solely extract the edge states to facilitate the calculation of the spectral winding number of the edge sates.  This is most obvious if we return to the spectrum shown in Fig.~\ref{Haldane}(e), where the energies of the edge states at the opposite sides of a strip structure are connected with delocalized states and hence there is no distinct closed spectral loops.  Physically, this is because the edge states at one side of the strip cannot be adiabatically connected with (or pumped to) those at the other side, unless there is a delocalization transition through the bulk states.

Following Refs. \cite{Kawabata2020a,Okugawa2020,Fu2021}, below we turn to an auxiliary Hamiltonian approach in order to reveal the nature of the found hybrid skin-topological modes induced by gain and loss in the Haldane model $H(\mathbf{k})$.  Specifically, let us consider the following Hamiltonian constructed from $H(\mathbf{k})$,
\begin{equation}\label{eq5}
  \tilde{H}(\mathbf{k},E_r)=\left(\begin{array}{cc}
                               0 & H(\mathbf{k})-E_r \\
                               H^\dag(\mathbf{k})-E_r & 0
                             \end{array}
  \right),
\end{equation}
with $E_r$ being a real number and representing a reference energy.   As seen above, the auxiliary Hamiltonian is Hermitian by construction.  Specifically, $\tilde{H}(\mathbf{k},E_r)$ lives on the expanded Hilbert space, the tensor product of the Hilbert space of $H(\mathbf{k})$ and that of an additional fictitious pseudo-spin degree of freedom, depicted by  another set of three Pauli matrices  $\tau_j$, $j=x,y,z$.    The above-defined auxiliary Hamiltonian can then be written as
\begin{equation}
 \tilde{H}(\mathbf{k},E_r) =\tau_+ (H(\mathbf{k})-E_r) +  \tau_- (H^\dag(\mathbf{k})-E_r)
\end{equation}
$\tau_{\pm}=\tau_x\pm\tau_y$. It is then seen clearly that the auxiliary Hermitian Hamiltonian satisfies the following chiral symmetry
\begin{equation}\label{eq6}
  \tau_z\tilde{H}(\mathbf{k},E_r)\tau_z=-\tilde{H}(\mathbf{k},E_r).
\end{equation}

The usefulness of the above-constructed auxiliary Hamiltonian can be appreciated as follows.  Consider 1D Hermitian systems with chiral symmetry, where the pseudo-spin degree of freedom is defined via $\tau_j$, $j=x,y,z$. The emergence of topological zero modes in such 1D systems is protected by a non-zero winding of vector $\{\Re[Q(k)]-E_r,\Im[Q(k)]\}$ around the origin of complex plane. Here $Q(k)=\det[H(\mathbf{k})]$ being the determinant of $H(\mathbf{k})$ is a complex number.   For example,  for the 1D Su-Schrieffer-Heeger (SSH) model $H_{\rm SSH}= h_x(k)\tau_x+ h_y(k)\tau_y$, the existence of topological zero modes can be directly connected, via the well-known bulk-edge correspondence, with the winding of the 2D vector $[h_x(k), h_y(k)]$ around the origin as the 1D momentum $k$ is scanned across the whole Brillouin zone.  With this preparation, let us now return to our 2D auxiliary model system  but with mixed boundary condition, namely, one direction is open and the second direction is under PBC, with $k_{\parallel}$ being the Bloch momentum parallel to the edge.   Let us further assume that the edge Hamiltonian of this 2D system is now reduced to
\begin{equation}
\tilde{H}_{\rm edge}(k_{\parallel},E_r) =\tau_+ (H_{\rm edge}(k_{\parallel})-E_r) +  \tau_- (H_{\rm edge}^\dag(k_{\parallel})-E_r),
\end{equation} which can be understood in parallel with a 1D model with chiral symmetry. Let $Q_{\rm edge}(k)=\det[H_{\rm edge}(k_{\parallel})]$.
One can then infer that the nontrivial winding of $\{\Re[Q_{\rm edge}(k_{\parallel})],\Im[Q_{\rm edge}(k_{\parallel})]\}$ around $E_r$ has a definite correspondence with the emergence of topological edge modes of the edge Hamiltonian (hence topological corner modes, if the $k_{\parallel}$ direction is also opened up).  Thus, the emergence of topological corner modes for the system $\tilde{H}(\mathbf{k},E_r)$ under OBC reflects the winding of $\{\Re[Q_{\rm edge}(k_{\parallel})],\Im[Q_{\rm edge}(k_{\parallel})]\}$ around $E_r$.  Remarkably, precisely it is the winding behavior of $\{\Re[Q_{\rm edge}(k_{\parallel})],\Im[Q_{\rm edge}(k_{\parallel})]\}$ around $E_r$ that one needs in order to confirm the occurrence of skin effects on the chiral edge states of the original non-Hermitian Chern insulator when the second direction is also placed under OBC.

\begin{figure}[htbp]
\includegraphics[width=1\linewidth]{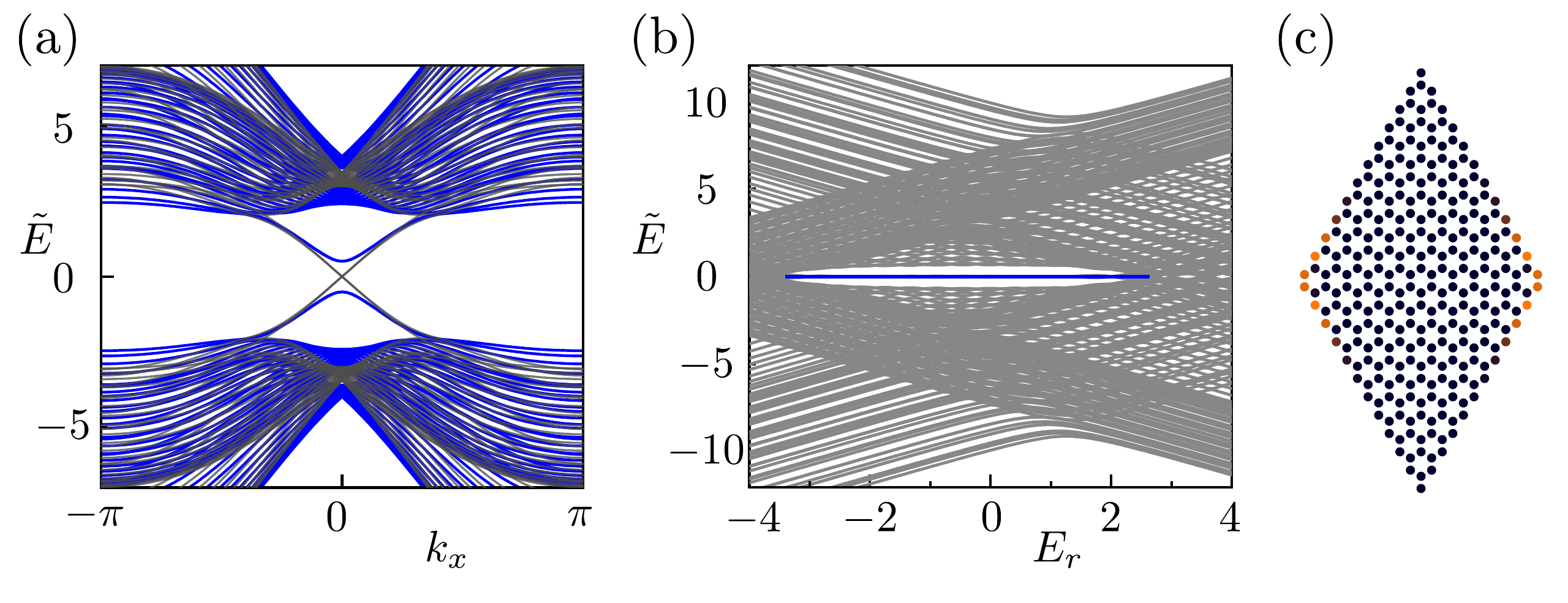}
\caption{(Color online) Second-order topological insulator of the auxiliary Hamiltonian defined in the main text.
(a) Energy bands of zigzag strip structure. Gray (blue) lines show the results without (with) gain and loss, $g=0$ ($g=0.6$).
(b) Spectrum of finite structure as a function of reference energy $E_r$.
(c) State density profile for the topological zero-energy corner states with $E_r=-t_2$.}
\label{HOTI}
\end{figure}

Insights above have made it clear the following bulk-edge-corner correspondence:  If $H_{\rm edge}(k_{\parallel})$ indeed has non-zero spectral winding number for the chiral edge states, or equivalently, if the auxiliary Hermitian Hamiltonian  $\tilde{H}(\mathbf{k},E_r)$ is actually a second-order topological insulator supporting topological corner modes, then under OBC in both directions, the non-Hermitian Chern insulator $H(\mathbf{k})$ hosts corner skin modes.
We are now ready to carry out a detailed analysis of the auxiliary Hamiltonian $\tilde{H}(\mathbf{k},E_r)$.
By substituting Eq.~(\ref{eq2}) into Eq.~(\ref{eq5}), the explicit auxiliary Hamiltonian for our non-Hermitian Haldane model reads as follows:
\begin{equation}\label{eq8}
\begin{split}
  \tilde{H}(\mathbf{k},E_r)=&[d_0(\mathbf{k})-E_r]\tau_x\sigma_0+d_x(\mathbf{k})\tau_x\sigma_x+ \\
  &d_y(\mathbf{k})\tau_x\sigma_y+d_z(\mathbf{k})\tau_x\sigma_z+g\tau_y\sigma_z. \\
\end{split}
\end{equation}
Here the second and third terms yield Dirac points at some high-symmetry momentum points.
Interestingly, the fourth term opens a band gap and gives rise to gapless edge states. The first term commutes with these three terms so that it just shifts the gapless edge state.
In the absence of any gain or loss, the auxiliary Hamiltonian hence supports gapless edge states.
The gray lines shown in Fig.~\ref{HOTI}(a) show one such example with $E_r=-t_2$.
In particular, the existence of gapless edge states clearly indicates that the auxiliary Hamiltonian without gain or loss is a first-order topological insulator.
Intriguing physics sets in when gain and loss are introduced.  It is seen from Fig.~\ref{HOTI}(a) (blue lines) that the edge states will acquire a band gap, hinting that the gain and loss introduced here, no matter how weak their strength is, can in fact induce a topological phase transition. Indeed,  the gain/loss term is reflected in the fifth term in Eq.~\ref{eq8}, which anti-commutes with the fourth term.   The gain/loss scheme advocated here is hence the main reason to induce gapped edge states, one main feature of second-order topological insulators.

To strengthen our findings on corner skin modes that can be obtained only under OBC for both dimensions, next we further confirm that the auxiliary Hermitian Hamiltonian $ \tilde{H}(\mathbf{k},E_r)$ is a second-order topological insulator by symmetry analysis and spectrum calculations.  Firstly, note that $H(\mathbf{k})$ supports pseudo inversion symmetry $\sigma_x H(\mathbf{k})\sigma_x=H^\dag(-\mathbf{k})$ so that the auxiliary Hamiltonian are inversion-symmetric, namely,  $I\tilde{H}(\mathbf{k},E_r)I^{-1}=\tilde{H}(-\mathbf{k},E_r)$ with $I=\tau_x\sigma_x$.
Upon a symmetry analysis applied to two lower bands at inversion symmetric momentum points, it is found that the auxiliary Hermitian Hamiltonian here does not have the Wannier representation.
However, it can be Wannierized by adding a trivial atomic insulator $s@q_{1c}$~\cite{Po2018} and is hence a fragile topological phase,  which can be expressed as $s@q_{1b}\oplus p@1_{1a}\oplus p@q_{1d}\ominus s@q_{1c}$ (See Appendix A).
Such fragile topological phase has half topological charge at the left and right corners.  Combined with the above-identified chiral symmetry,  this auxiliary system can be predicted to support topological zero-energy corner modes~\cite{Benalcazar2019}.  Fig.~\ref{HOTI}(b) presents the spectrum of the auxiliary system with OBC in both directions, as a function of $E_r$.
The auxiliary system is seen to support topological zero-energy states in the range of $-3.5\lesssim E_r\lesssim 2.5$, which agrees perfectly with the spectrum of the skin corner modes of our non-Hermitian Chern insulator under OBC,  as shown in Fig.~\ref{Haldane}(f).   Fig.~\ref{HOTI}(c) further presents state density profiles of the found topological zero-energy states of $ \tilde{H}(\mathbf{k},E_r)$, with $E_r=-t_2$.  These topological states are indeed localized at the corners and hence are second-order topological corner states.
This comparison in terms of the zero-energy states between the auxiliary system and our actual non-Hermitian Chern insulator
verifies our physical insights above.

\begin{figure}[htbp]
\includegraphics[width=\linewidth]{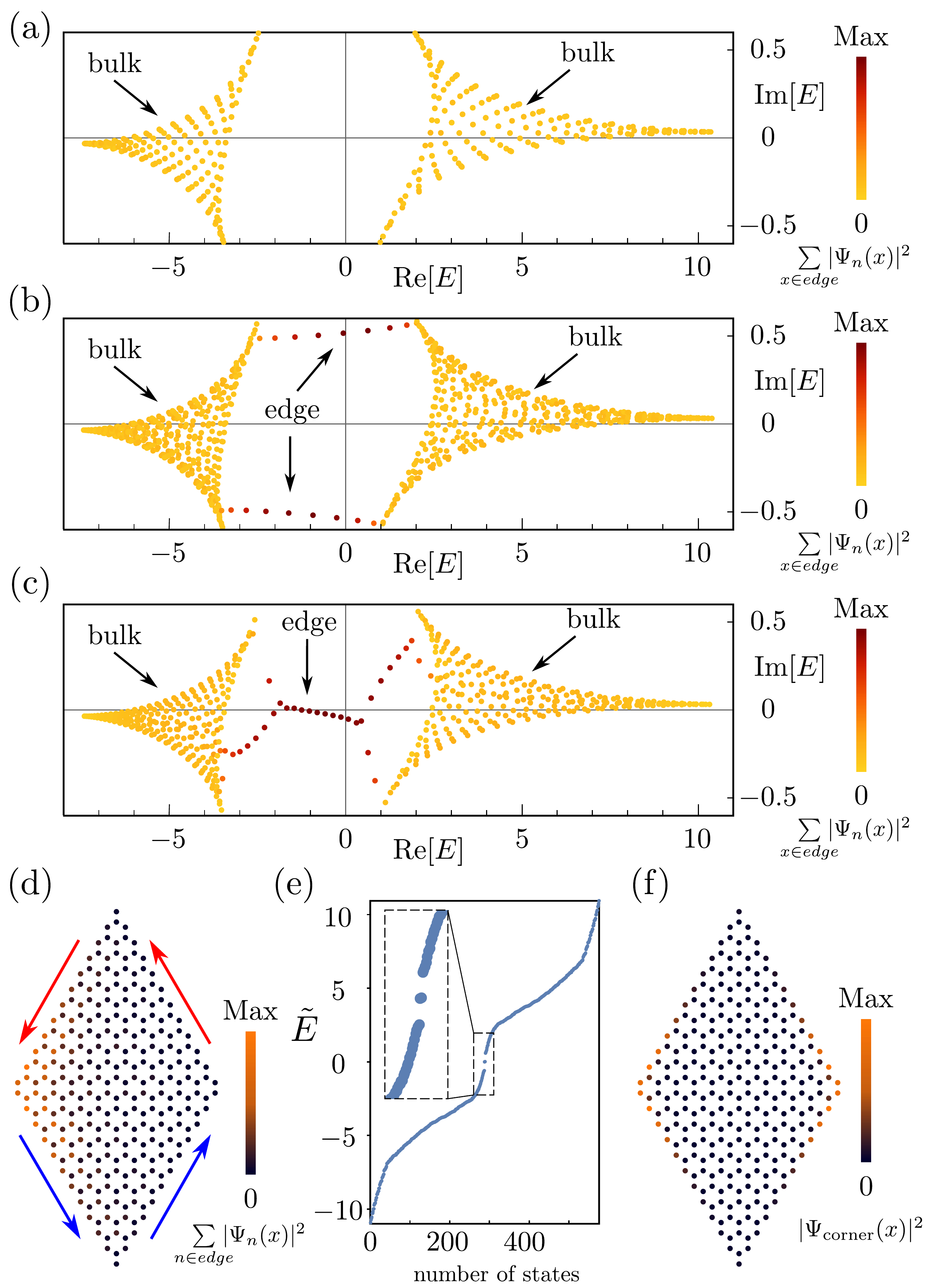}
\caption{Spectrum and hybrid skin-topological modes in a non-Hermitian Haldane model with on-site potential difference introduced. (a)-(c): Spectrum of the system under PBC, mixed PBC-OBC, and OBC, respectively.  Color represents the strength of localization on the left edge $\sum_{x\in edge}|\Psi_n(x)|^2$.
(d) Summation of state densities for all edge states $\sum_{n\in edge}|\Psi_n(x)|^2$.
(e) Spectrum of the constructed auxiliary Hamiltonian. (f) The profile of zero-energy topological corner mode shown in (e) In all the calculations, $\mu$ is chosen as 0.5.}
\label{onsite}
\end{figure}

Our discussions so far have connected with the hybrid skin-topological modes with an auxiliary Hamiltonian as a second-order topological insulator protected by inversion symmetry and chiral symmetry.  The involved second-order topological insulator phase is of the intrinsic type, because this phase is protected by the bulk topology with crystal symmetry, and so no topological phase transitions can occur without bulk-gap closing or symmetry breaking.  This brings us an interesting question as follows: is it possible to have hybrid skin-topological modes with a second-order topological insulator of the extrinsic type, where the second-order topological phase is only protected by a gap of the edge states or by the edge topology~\cite{Geier2018}.

To address this interesting question we now introduce a weak on-site potential difference ($\mu\sum_{i\in A}c_i^\dag c_i-\mu\sum_{i\in B}c_i^\dag c_i$) to Eq.~(\ref{eq1}).  This additional term breaks the pseudo inversion symmetry of $H(\mathbf{k})$ so that it also breaks the inversion symmetry of the associated auxiliary Hamiltonian.  Without such crystal symmetry, the obtained second-order topological insulator phase is not protected by a bulk gap.

Consider then first the spectrum of the system $H+ \mu\sum_{i\in A}c_i^\dag c_i-\mu\sum_{i\in B}c_i^\dag c_i$, under PBC, mixed PBC-OBC, and OBC, as shown in Fig.~\ref{onsite}(a)-(c).  The main spectral features observed in the previous case are also observed here.   Fig.~\ref{onsite}(a) depicts two bulk bands under PBC. Under mixed PBC-OBC in Fig.~\ref{onsite}(b),  edge states are observed on top of the bulk spectrum, with the whole spectrum enclosing a nonzero area.  The energetics of the edge states change dramatically again when we change from mixed PBC-OBC to OBC in Fig.~\ref{onsite}(c).  Interestingly, regardless of the boundary condition, the bulk states are always extended.  Fig.~\ref{onsite}(d) shows that the edge states are localized at one corner when the system is under OBC.

\begin{figure}[htbp]
\includegraphics[width=\linewidth]{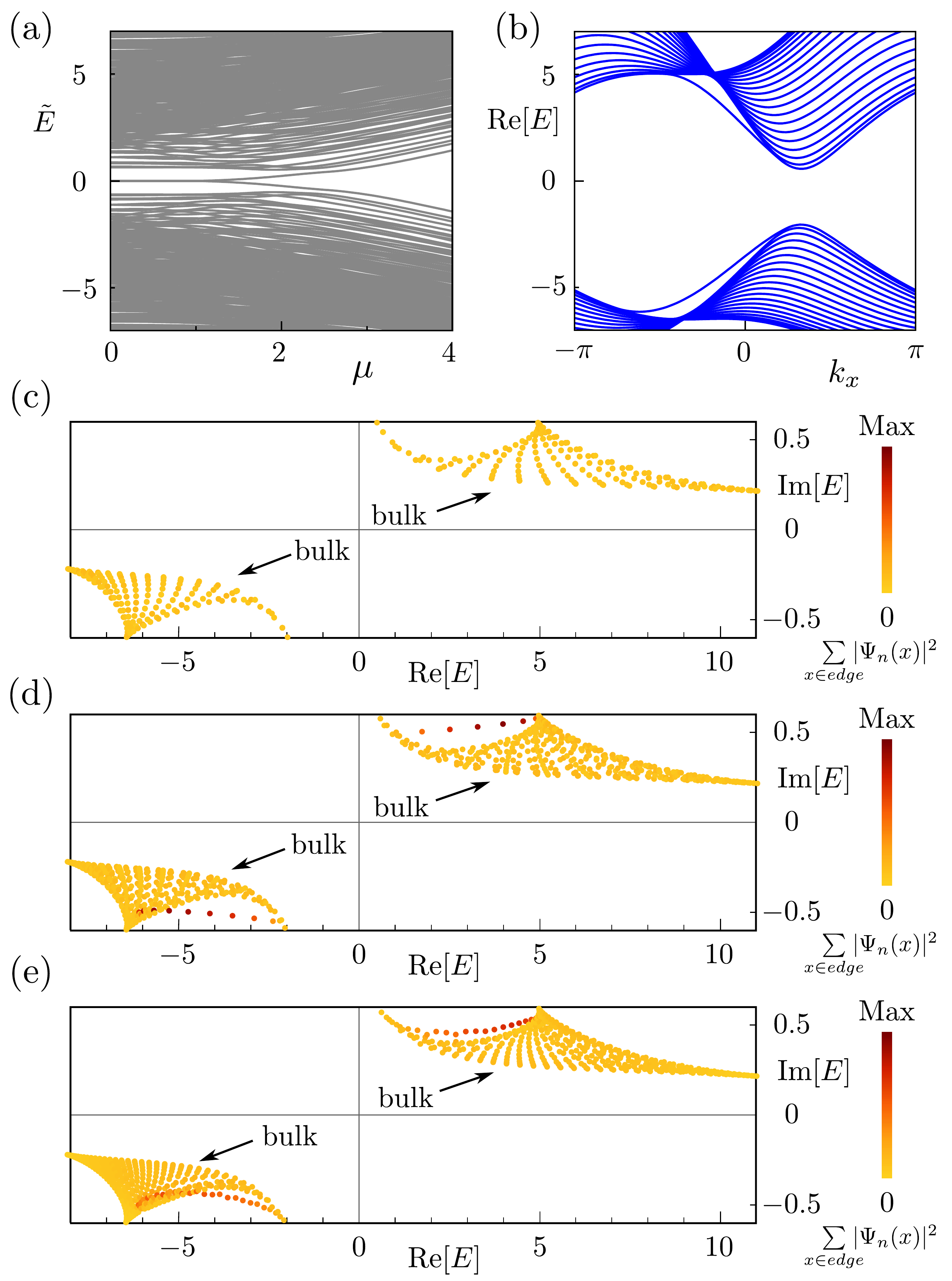}
\caption{Topological phase transition causing hybrid skin-topological modes to disappear, thus offering a topological switch to control skin effects.  (a) Spectrum of the auxiliary Hamiltonian associated with our non-Hermitian Haldane model, where an on-site potential difference with a varying strength parameter $\mu$ is introduced, with fixed $E_r=0$. (b) Real part of the energy bands for our non-Hermitian Haldane model vs Bloch momentum $k_x$ under mixed PBC-OBC.  (c)-(e): Spectrum of the system plotted on the complex energy plane, under PBC, mixed PBC-OBC, and OBC, respectively.  Color represents the strength of localization on the left edge $\sum_{x\in edge}|\Psi_n(x)|^2$. $\mu$ is chosen as $3.5$ for panels (b)-(e).}
\label{switch}
\end{figure}

To verify that these states localized at corner arise from NHSE, we investigate in parallel the spectrum and localization behavior of
the corresponding auxiliary Hamiltonian.   The spectrum of the  auxiliary Hamiltonian under OBC is shown in Fig.~\ref{onsite}(e), with in-gap states observed. However, due to the lack of the inversion symmetry possessed in the previous case,
the in-gap states are only protected by a gap in the edge states, not by the bulk gap.  That is, the presence of edge states is not in one-to-one correspondence with bulk gap closure/opening.  The in-gap states can thus only be interpreted as the edge states of 1D edge states and hence the auxiliary Hamiltonian is an extrinsic second-order topological insulator.
Nevertheless, Fig.~\ref{onsite}(f) verifies that this auxiliary Hamiltonian does support second-order corner modes.
The agreement is also checked quantitatively, insofar as the auxiliary Hamiltonian only supports topological zero-energy states in the range of $-3.2\lesssim E_r\lesssim 2.2$ (See Appendix B), fully consistent with the energetics of the corner states shown in Fig.~\ref{onsite}(c).

The hybrid skin-topological corner modes revealed in this work suggests that we may envision the construction of a topological switch to turn on/off the skin effect via topological phase transitions.  To explore this possibility, we tune the above-introduced  on-site potential difference $\mu$ to induce a topological phase transition, from a Chern insulator to a normal insulator.  With this topological phase transition,  the underlying mechanism for hybrid skin-topological corner modes ceases to exist and hence the skin effect should disappear as well.    Fig.~\ref{switch}(a) depicts this topological  phase transition via the spectrum of the auxiliary Hamiltonian.
Specifically, the auxiliary systems changes from a second-order topological insulator to a normal insulator with the increase of $\mu$, with the topological phase transition point at around $\mu\approx2$.   Taking the case of $\mu=3.5$ as an example,  the non-Hermitian Haldane model introduced here features a normal insulator, as shown in Fig.~\ref{switch}(b) with no gapless topological chiral edge states.
The overall spectrum under PBC, mixed PBC-OBC, and OBC of our system is shown in Fig.~\ref{switch}(c)-(e) on the complex plane. It is now seen that not only the bulk states are insensitive to the boundary conditions, but also the edge states become insensitive to the boundary conditions.  Consistent with these observations, hybrid skin-topological modes are not obtained.    This strengthens our understanding of the newly discovered class of hybrid skin-topological modes. Indeed,   the edge states seen in  Fig.~\ref{switch}(d)-(e) are just defect states and not chiral (unidirectional), and hence there will not be net accumulation of gain or loss for these defect states. 

\section{Nonequilibrium hybrid skin-topological modes}
\subsection{Hybrid skin-topological modes in a non-Hermitian periodically driven model}
\begin{figure}[htbp]
\includegraphics[width=1\linewidth]{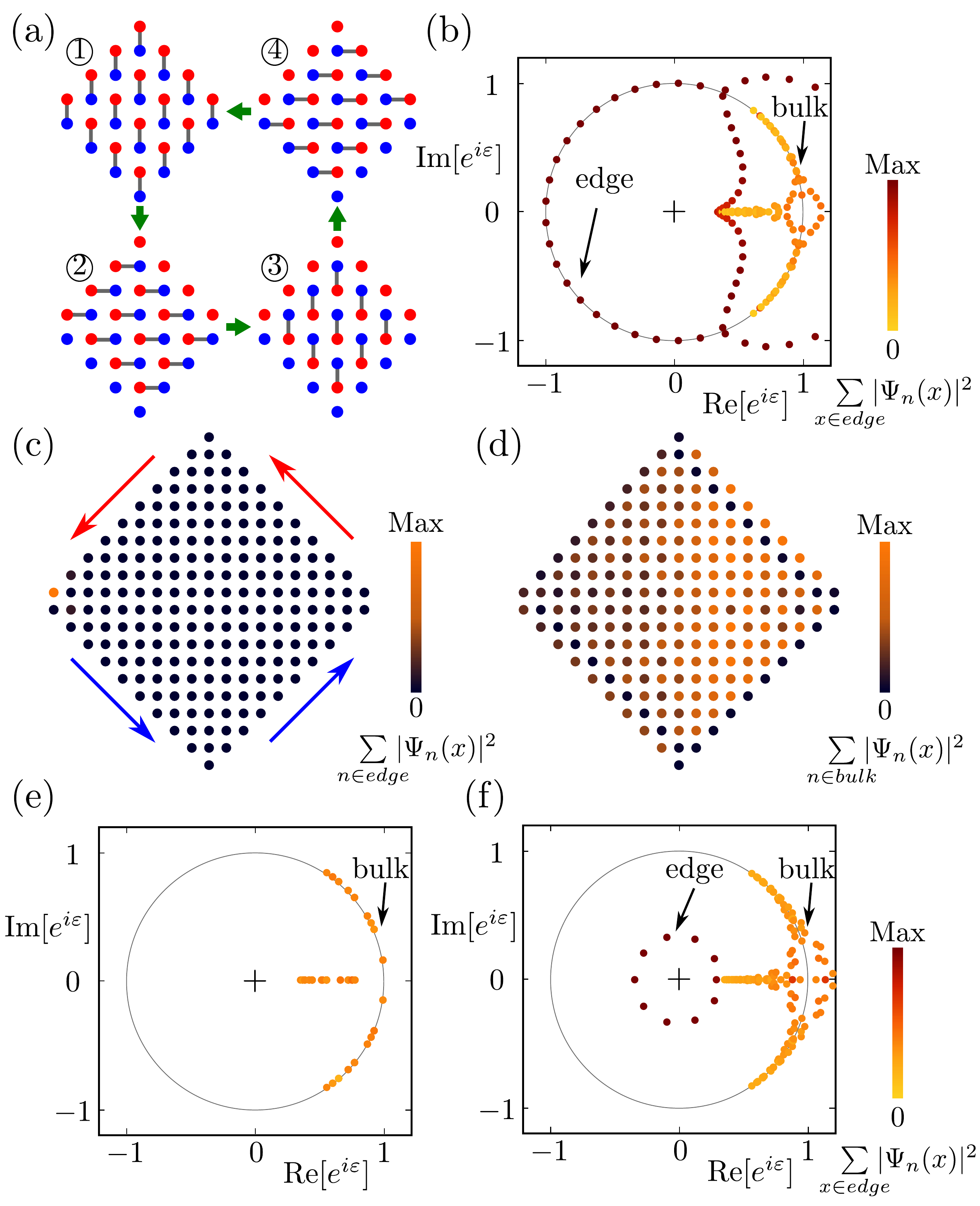}
\caption{(Color online) Hybrid skin-topological modes in a non-Hermitian Floquet system with chiral edge states.
(a) A periodically driven lattice model proposed in Ref.~\cite{Rudner2013} with gain and loss introduced to two sublattices.
(b) Quasi-energies plotted on the complex plane of $e^{i\varepsilon}$ when the system is under OBC.  The unit circle is marked where $\varepsilon$ is real. Color represents the strength of localization on the left edge $\sum_{x\in edge}|\Psi_n(x)|^2$.
(c) Summation of state densities for all edge states $\sum_{n\in edge}|\Psi_n(x)|^2$.
(d) Summation of state densities for all bulk states $\sum_{n\in bulk}|\Psi_n(x)|^2$. (e) Same as in panel (b) but under PBC.  (f) Same as in panel (b) but the system is under mixed PBC-OBC.
The system parameters chosen are $\theta=0.6\pi$ and $g=0.5$.}
\label{floquet}
\end{figure}
To further confirm that our route towards the hybrid skin-topological modes is widely applicable,  we now apply the same strategy to periodically driven (Floquet) topological phases~\cite{Kitagawa2010,Ho2012,Gomez-Leon2013,Kundu2013,Lababidi2014,FoaTorres2014,Hubener2017,Zhou2018,McIver2019,Rudner2020,Wintersperger2020}.  Given that nonequilibrium topological matter is less well understood due to the possibility of anomalous chiral edge states.  It is indeed timely to investigate the consequences of adding gain and loss to Floquet topological matter.  To our knowledge, so far there is no study whatsoever on hybrid skin-topological modes in nonequilibrium systems.

Without loss of generality, let us consider a model of anomalous Floquet topological phase,  as first proposed by in Ref.~\cite{Rudner2013}. This model can be realized with an explicit platform, i.e., coupled ring resonators~\cite{Pasek2014,Zhu2021a,Zhu2021b}.
The time-dependent Bloch Hamiltonian consists of four steps,
\begin{eqnarray}
 H(\mathbf{k},t)=\left\{
\begin{array}{cc}
H_{1}(\mathbf{k})&0<t\leq T/4\;,\\
H_2(\mathbf{k})&T/4<t\leq T/2\;,\\
H_3(\mathbf{k})&T/2<t\leq 3T/4\;,\\
H_4(\mathbf{k})&3T/4<t\leq T\;,
\end{array}
\right.
\label{eq9}
\end{eqnarray}
where
\begin{equation}
H_{m}(\mathbf{k})=4*\theta(e^{ib_m\cdot\mathbf{k}}\sigma^{+}+h.c.)+ig\sigma_z,
\label{eq10}
\end{equation}
for $m=1,2,3,4$. Here $\sigma^{+}=(\sigma_x+i\sigma_y)/2$ and the vectors $\mathbf{b}_{m}$ are given by $\mathbf{b}_{1}=(0,0)$, $\mathbf{b}_{2}=(1/\sqrt{2},1/\sqrt{2})$, $\mathbf{b}_{3}=(0,\sqrt{2})$ and $\mathbf{b}_{4}=(-1/\sqrt{2},1/\sqrt{2})$. We set $T=1$. As also seen above, the gain and loss are also introduced to two sublattice through the term $ig\sigma_z$.  Fig.~\ref{floquet}(a)  illustrates this nonequilibrium topological phase model.

The quasi-energy Floquet bands $\varepsilon(\mathbf{k})$ of the above-described nonequilibrium system can be obtained by solving the following eigen-equation,
\begin{equation}\label{eq11}
  U_T(\mathbf{k})\, |\Psi(\mathbf{k})\rangle = e^{-i\varepsilon(\mathbf{k})T} |\Psi(\mathbf{k})\rangle\;,
\end{equation}
where $U_T(\mathbf{k}) \equiv \mathcal{T}\mathrm{exp}\big[\!-\!i\!\int_{0}^{T} H(\mathbf{k},\tau)\,d\tau\big]$ is the Floquet operator, $\mathcal{T}$ is the time-ordering operator.

Let us first examine the spectrum for both dimensions under OBC.  In our calculations, we  set $\theta=0.6\pi$ as an example.  For this choice,  the system with $g=0$ supports chiral edge states~\cite{Zhu2021a}.  With gain and loss at a small strength switched on, the chiral edge states persist.  The results for $g=0.5$ are shown in Fig.~\ref{floquet}(b), where a single bulk band and gapless topological edge states are observed.
All the edge states are localized at the left corner of the 2D lattice, as shown in Fig.~\ref{floquet}(c). By contrast,  the bulk states are still extended as shown
in Fig.~\ref{floquet}(d).  All these features are analogous to the previous non-Hermitian Chern insulator case and hence one can expect that hybrid skin-topological modes can be also induced here.
 In  Fig.~\ref{floquet}(e) and   Fig.~\ref{floquet}(f) we also present the spectrum when the system is under PBC and mixed PBC-OBC, respectively.   Again, the energetics of the nonequilibrium chiral edge states are sensitive to the boundary condition, with the PBC-OBC case featuring one clear loop in the spectrum (here we only show the edge states with gain, the edge states with loss out the unit circle also forms a loop).

\subsection{An auxiliary Hamiltonian approach to nonequilibrium cases}
To investigate on a solid ground whether the above-observed corner modes are hybrid skin topological modes, it is necessary to examine  the winding behavior of the energetics of the Floquet chiral edge states, again through the emergence of topological corner modes of an auxiliary Hamiltonian.   For Floquet systems, it is tempting to define a Floquet Hamiltonian $H_F(\mathbf{k})=(i/T)\ln(U_T(\mathbf{k}))$, whose eigenvalues are connected with the quasi energies $\varepsilon(\mathbf{k})$. In this naive approach, an auxiliary Hamiltonian analogous to Eq.~(\ref{eq5}) can be defined from $H_F(\mathbf{k})$.
However, such definition of a Floquet Hamiltonian $H_F(\mathbf{k})$ actually needs to pre-define branch cuts for different quasi-energy gaps in order to take the logarithm operation without ambiguity.  In addition, the full topology of a Floquet topological system cannot be captured solely by $H_F(\mathbf{k})$, given that here the Floquet band Chern numbers are zero but there are still chiral edge states~\cite{Zhu2021b}.   As one contribution of this work, we avoid involving  $H_F(\mathbf{k})$ and propose to examine instead the winding behavior of $U_T(\mathbf{k})$.

Specifically, let us now construct the following auxiliary Hamiltonian with the Floquet operator $U_T(\mathbf{k})$ directly,
\begin{equation}\label{eq12}
  \tilde{H}(\mathbf{k},\varepsilon_r)=\left(\begin{array}{cc}
                               0 & U_T(\mathbf{k})-e^{-i\varepsilon_r} \\
                               U_T^\dag(\mathbf{k})-e^{i\varepsilon_r} & 0
                             \end{array}
  \right),
\end{equation}
where $\varepsilon_r$ is a real number ranging from $0$ to $2\pi$ as a reference quasi-energy.   By construction, this auxiliary Hamiltonian is Hermitian.   Analogous to our previous construction, this auxiliary system also possess the chiral symmetry. The existence of topological corner modes indicates the nontrivial winding of the edge Hamiltonian as $k_{\parallel}$, the Bloch momentum parallel to the edge, varies over one period.  This then suggest the nontrivial winding of the quasi-energy of $U_T(\mathbf{k})$ if one dimension is under OBC and hence the skin effect on the chiral edge states.   Note also that our concern here is not about predicting whether there are chiral edge states through a full topological characterization and only spectral winding is important, so the use of   $U_T^\dag(\mathbf{k})$ above suffices to understand the consequences of gain and loss on any existing chiral edge states.

\begin{figure}[htbp]
\includegraphics[width=1\linewidth]{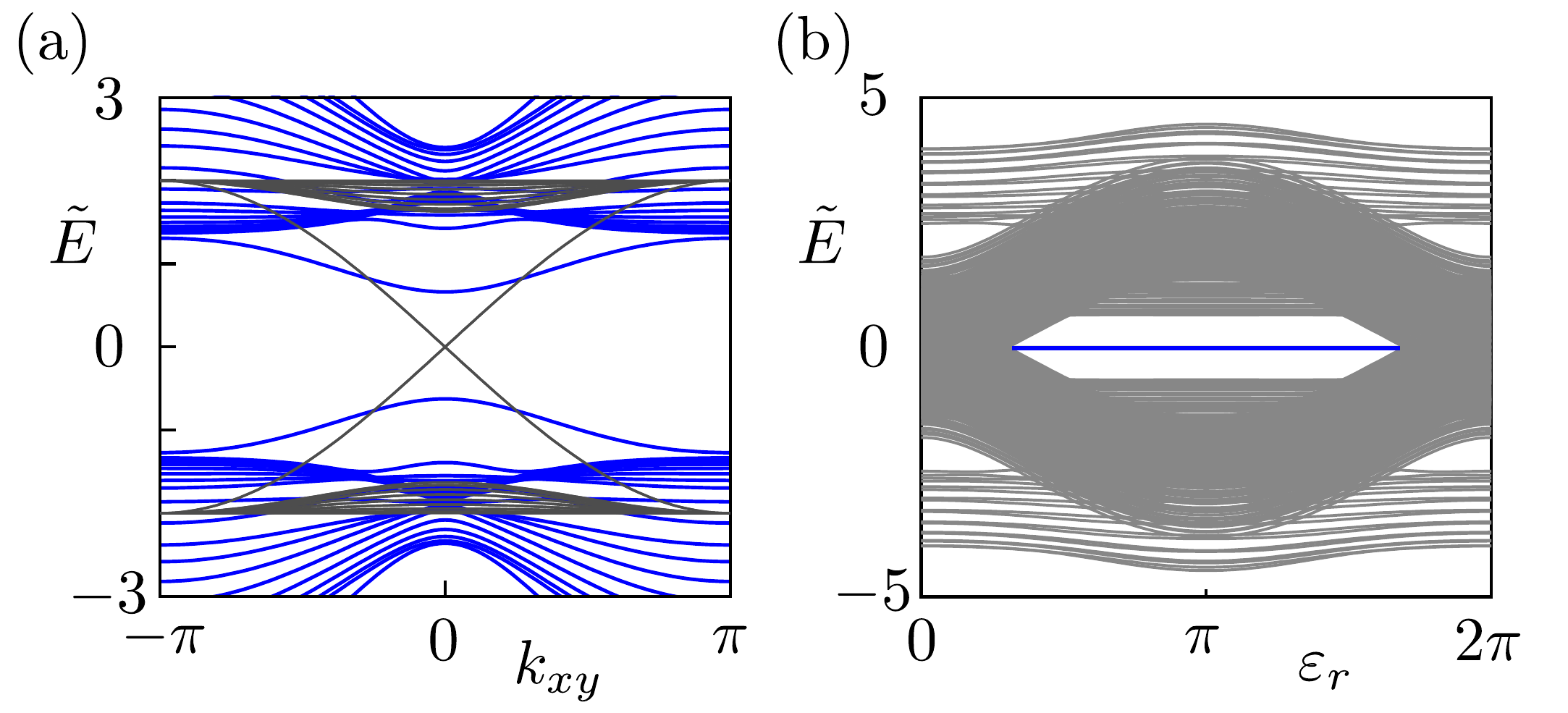}
\caption{(Color online) Second-order topological insulator of the auxiliary Hamiltonian associated with the non-Hermitian Floquet model.
(a) Energy bands of strip structure for $\varepsilon_r=\pi$. Gray (blue) lines show the results without (with) gain and loss, $g=0$ ($g=0.5$).
(b) Spectrum of finite structure as a function of reference quasi-energy $\varepsilon_r$.
The parameters we used are $\theta=0.6\pi$ and $g=0.5$.}
\label{fedge}
\end{figure}
Fig.~\ref{fedge}(a) presents the energy bands of  $\tilde{H}(\mathbf{k},\varepsilon_r)$  under a strip structure (mixed PBC-OBC).
The edge states are gapless without gain and loss ($g=0$), as shown by the grey lines.  However,
upon introducing gain and loss, the edge states of the auxiliary Hamiltonian opens a band gap,  as shown by the blue lines.  This feature is similar to the previous case constructed from the non-Hermitian Chern insulator, where the gain and loss have induced a  topological transition from first-order insulator to second-order insulator. Indeed, if both dimensions are under OBC,  this auxiliary system supports topological corner states, as shown in Fig.~\ref{fedge}(b).
In particular, if the quasi-energy parameter $\varepsilon_r$ is in the range of $[0.3\pi$, $1.7\pi]$, second-order zero-energy corner states are obtained.
This range agrees precisely with that of the quasi-energy of skin corner states plotted on the unit circle in Fig.~\ref{floquet}(b).
We have thus computationally and conceptually demonstrated that even in nonequilibrium topological systems, the occurrence of hybrid skin-topological modes, which correspond to the winding behavior of the quasi-energy of the edge states when one dimension is under OBC,  comes in parallel with second-order topological insulator phase of an auxiliary Hamiltonian constructed directly from the Floquet operator.  Our approach here is expected to be widely useful when inspecting the existence of nonequilibrium hybrid skin-topological modes.

\section{Discussion}
It is made clear above that the hybrid skin-topological modes arise from the skin effect localization of topological chiral edge states, in both the static and the periodically driven models as case studies. We highlight the similarity and difference between our results and the previous results of hybrid skin-topological modes~\cite{Lee2019,Li2020a,Zou2021,Shang2022}.
In both cases, the number of corner skin modes is proportional to the length of the system and the corner localization appears only if the system is under OBC for two dimensions. However, in previous examples of skin-topological modes,  the role played by two dimensions is entirely separate, one for topological localization and one for skin localization.    By contrast,  in our generic route here,  the two system dimensions are under the same footing, and the obtained skin-topological corner modes are deeply connected with some second-order topological phase induced by gain and loss. Our system is simultaneously a non-Hermitian Chern insulator/non-Hermitian anomalous Floquet topological insulator and supports gapless topological edge state.  Furthermore, our proposal is so general that it is extendable to nonequilibrium situations, leading us to the finding of nonequilibrium skin-topological modes.

The two working models considered in this work are highly feasible for experimental studies.
Indeed, their Hermitian counterparts (without gain or loss) have already been realized in photonics and acoustics~\cite{Rechtsman2013,Gao2016,Peng2016,Mukherjee2017,Maczewsky2017,Mittal2019a,Afzal2020,Zhu2022}.
Although in our theoretical considerations we introduce both gain and loss to the system, in real experiments only loss suffices because only the difference between two sublattice sites matters. The loss can be realized by sound/optical absorbing materials or leaky modes~\cite{Ding2016,Noh2017,Xiao2017,Zhu2018,Ozdemir2019,Gao2021}.  In future, it is also interesting to study the discovered mechanism  in continuous systems, like topological photonic/phononic crystal~\cite{Ozawa2019,Ma2019}, instead of lattice models.
Considering the seminal Kane-Mele model can be treated as two copies of the Haldane model, it is also possible to find gain and loss induced second-order NHSE in the Kane-Mele model or other  valley topological insulators and topological crystal insulators~\cite{Qi2011}.

\bigskip


\vspace{0.3cm}
{\bf Acknowledgements:}
J.G. acknowledges fund support by the Singapore Ministry of Education Academic
Research Fund Tier-3 Grant No. MOE2017-T3-1-001
(WBS. No. R-144-000-425-592) and by the Singapore National Research Foundation
Grant No. NRF-NRFI2017- 04 (WBS No. R-144-000-378-
281).  We thank Ching Hua Lee and Linhu Li for very helpful discussions.


\begin{appendix}




\section{Appendix A: Symmetry analysis of an auxiliary Hamiltonian associated with the non-Hermitian Haldane model.}

In this section, we present our finding that the auxiliary Hamiltonian of our non-Hermitian Haldane model, as defined in the main text, is in fact a fragile topological insulator with fractional charge at its corners.
Together with the chiral symmetry  of the auxiliary Hamiltonian by construction,  this system supports topological corner modes.

\begin{figure}[htbp]
\includegraphics[width=0.9\linewidth]{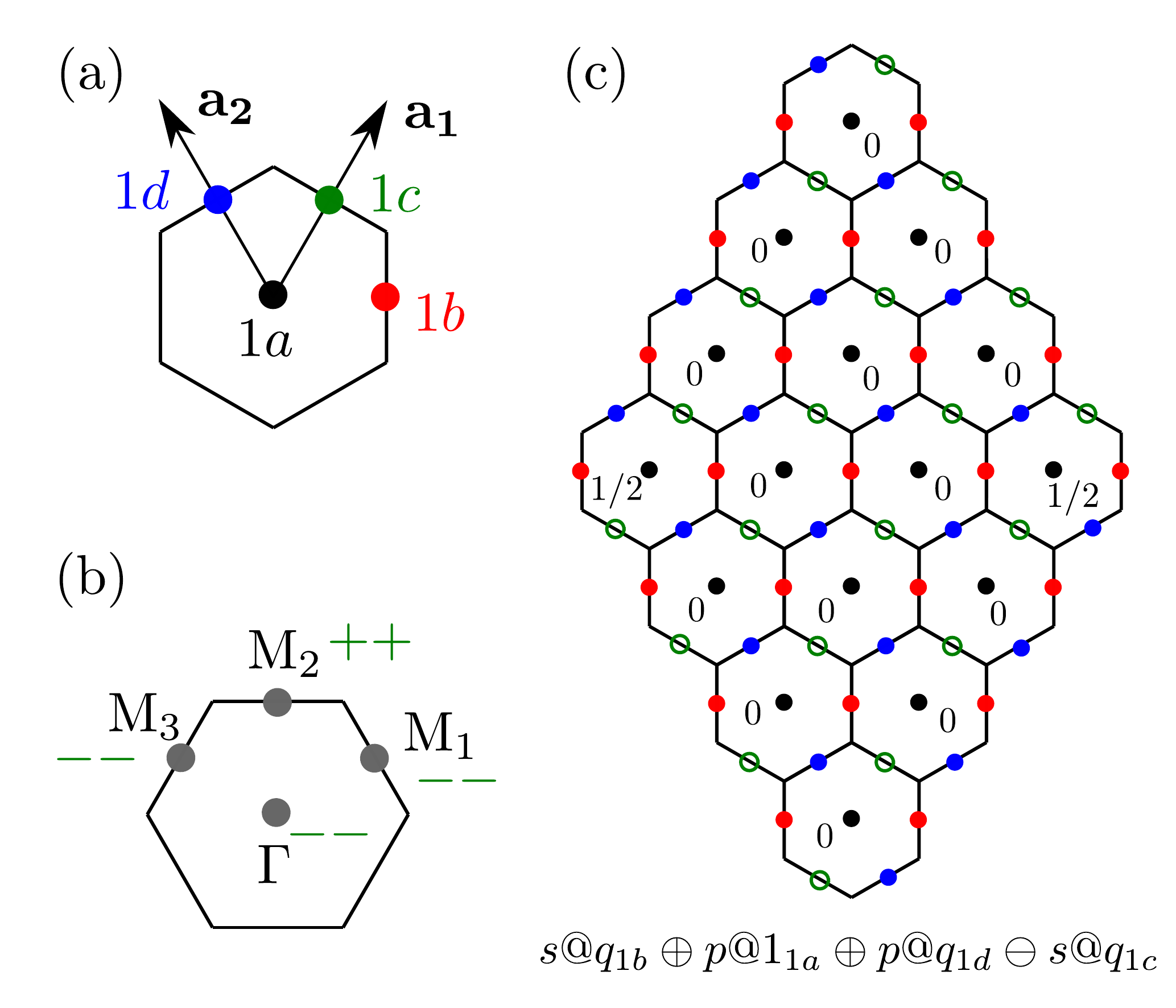}
\caption{Fractional charge for the auxiliary Hamiltonian of our non-Hermitian Haldane model. (a) The unit cell of the honeycomb lattice. The inversion symmetry points are marked as $1a$, $1b$, $1c$ and $1d$. (b) The first Brillouin zone. The inversion symmetry momentum points are marked $\Gamma$, $\mathrm{M}_1$, $\mathrm{M}_2$ and $\mathrm{M}_3$. The symmetry eigenvalues for the auxiliary Hamiltonian of non-Hermitian Haldane model are marked. (c) Two electrons with representation $s@q_{1b}\oplus p@1_{1a}\oplus p@q_{1d}\ominus s@q_{1c}$. The electron charges at different unit cells are indicated mod 1 (in units of the electron charge $e$).}
\label{symmetry}
\end{figure}

In Fig.~\ref{symmetry}(a), we show the unit cell of the honeycomb lattice and its inversion symmetry points $1a$, $1b$, $1c$ and $1d$.
The first Brillouin zone and the inversion symmetric momentum points $\Gamma$, $\mathrm{M}_1$, $\mathrm{M}_2$, $\mathrm{M}_3$ are shown in Fig.~\ref{symmetry}(b).
The elementary band representations of the system can be obtained by putting $s$ or $p$ orbital to the inversion symmetry points and are described by the symmetry eigenvalues at high symmetry momentum points~\cite{Cano2021}.
The elementary band representations are summarized in Table.\ref{table}.

It is seen that there are two bands below the band gap of the auxiliary Hamiltonian.
The symmetry eigenvalues at high symmetry momentum points for $t_1=3$, $t_2=0.5$, $g=0.6$ and $\phi=\pi/3$ are shown in Fig.~\ref{symmetry}(b).
Compared with the elementary band representations, it is seen that there is no Wannier representation for this auxiliary Hamiltonian.
However, we can add one atom orbital $s@q_{1c}$ to the two bands, then it can be represented as three atomic orbital $s@q_{1b}\oplus p@1_{1a}\oplus p@q_{1d}$.
This can then be identified as a fragile topological insulator phase, which does not have its Wannier representation but can be Wannerized by adding one atomic orbital.
The two bands then can be represented as $s@q_{1b}\oplus p@1_{1a}\oplus p@q_{1d}\ominus s@q_{1c}$.

\begin{table}[htbp]
\caption{\label{table}Inversion symmetry eigenvalues at high symmetry moment points for eight elementary band representations of honeycomb lattice with inversion symmetry. Each band representation has either 0, 2 or 4 negative eigenvalues. }
\begin{ruledtabular}
\begin{tabular}{ccccc}
& $\,\Gamma\,$ & $\,$ $\mathrm{M}_1$ $\,$& $\,$ $\mathrm{M}_2$ $\,$ & $\,$ $\mathrm{M}_3$ $\,$ \\
\hline
  $s@q_{1a}$ & $+1$ & $+1$ & $+1$ & $+1$ \\
  $p@q_{1a}$ & $-1$ & $-1$ & $-1$ & $-1$ \\
  $s@q_{1b}$ & $+1$ & $-1$ & $+1$ & $-1$ \\
  $p@q_{1b}$ & $-1$ & $+1$ & $-1$ & $+1$ \\
  $s@q_{1c}$ & $+1$ & $-1$ & $-1$ & $+1$ \\
  $p@q_{1c}$ & $-1$ & $+1$ & $+1$ & $-1$ \\
  $s@q_{1d}$ & $+1$ & $+1$ & $-1$ & $-1$ \\
  $p@q_{1d}$ & $-1$ & $-1$ & $+1$ & $+1$ \\
\end{tabular}
\end{ruledtabular}
\end{table}

In Fig.~\ref{symmetry}(c), we show the charge distribution for the two bands if the system is placed under OBC.
The solid circles with color black, blue and red means one orbital at $1a$, $1d$ and $1b$, respectively.
The open circle with color green means removing one orbital at $1c$.
By numbering these sites that keep the inversion symmetry, we can see the fractional charge for the left corner and right corner is $1/2$.
Together with the chiral symmetry, the system supports topological corner modes~\cite{Benalcazar2019}.

\section{Appendix B: Spectrum of the auxiliary Hamiltonian for a non-Hermitian Haldane model with on-site potential difference.}

\begin{figure}[htbp]
\includegraphics[width=\linewidth]{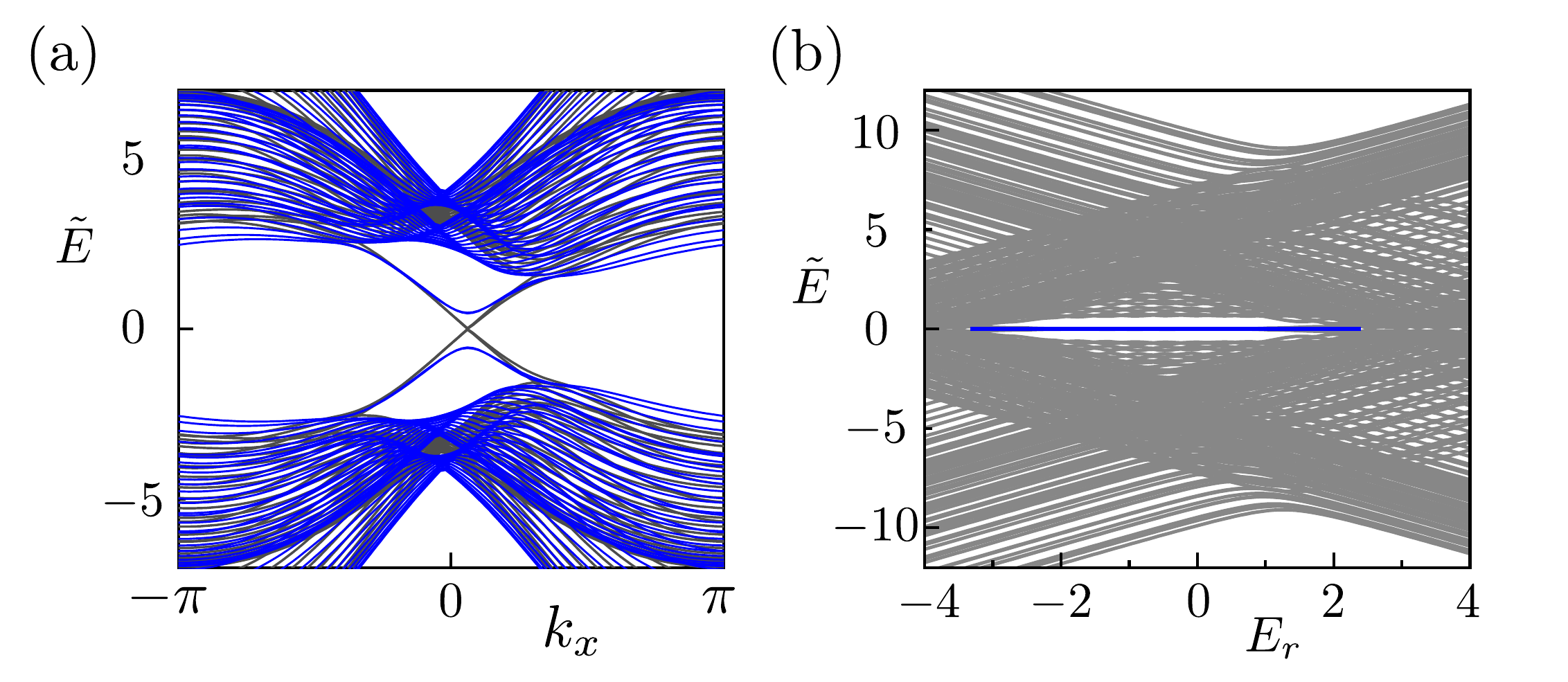}
\caption{ Second-order topological insulator of the auxiliary Hamiltonian for the non-Hermitian Haldane model with on-site potential difference $\mu\sum_{i\in A}c_i^\dag c_i-\mu\sum_{i\in B}c_i^\dag c_i$.
(a) Energy bands of zigzag strip structure. Gray (blue) lines show the results without (with) gain and loss, $g=0$ ($g=0.6$). The reference energy is $E_r=-t_2$.
(b) Spectrum of a finite structure as a function of the reference energy $E_r$. We choose $t_1=3$, $t_2=0.5$, $g=0.6$, $\phi=\pi/3$ and $\mu=0.5$.}
\label{extrinsic}
\end{figure}

In Fig.~4 of the main text, we already presented the main spectral results for the non-Hermitian Haldane model with on-site potential difference $\mu\sum_{i\in A}c_i^\dag c_i-\mu\sum_{i\in B}c_i^\dag c_i$.
Here we present more results complementary to Fig.~4.
The energy bands of the auxiliary Hamiltonian in zigzag strip structure with $E_r=-t_2$ are shown in Fig.~\ref{extrinsic}(a).
When there is no gain/loss, the system supports gapless edge state (grey lines). And the introduction of gain/loss open a band gap to the edge state (blue lines).
Due to on-site potential breaking the inversion symmetry, here the spectrum is not symmetric anymore.
The spectrum of the auxiliary Hamiltonian under OBC as a function of the reference energy $E_r$ is shown in Fig.~\ref{extrinsic}(b).
The system is seen to support zero corner mode for the range $-3.2\lesssim E_r\lesssim 2.2$,  which is fully consistent with the band gap of Fig.~4(a) shown in the main text.

\end{appendix}


%

\end{document}